\documentclass[10pt, onecolumn]{article}   	
\usepackage{geometry}                		
\pdfoutput=1                 		
\usepackage{graphicx}
\usepackage{subfigure}				
\usepackage{amssymb}
\usepackage[misc,geometry]{ifsym}
\usepackage{url}
\title{\bf Analysis of the Human-Computer Interaction on the Example of Image-based CAPTCHA by Association Rule Mining}
\author{Darko Brodi\'c$^{1(\mbox{\scriptsize\Letter})}$, Alessia Amelio$^2$}
\date{\small $^1$ Technical Faculty in Bor, University of Belgrade, Vojske Jugoslavije 12, 19210 Bor, Serbia, \\dbrodic@tfbor.bg.ac.rs\\$^2$ Department of Computer Engineering, Modeling, Electronics and Systems, University of Calabria, \\Via P. Bucci Cube 44, 87036 Rende (CS), Italy,\\aamelio@dimes.unical.it}							

\begin{document}
\maketitle

\begin{@twocolumnfalse}

{\bf Abstract}.
The paper analyzes the interaction between humans and computers in terms of response time in solving the image-based CAPTCHA. In particular, the analysis focuses on the attitude of the different Internet users in easily solving four different types of image-based CAPTCHAs which include facial expressions like: animated character, old woman, surprised face, worried face. To pursue this goal, an experiment is realized involving 100 Internet users in solving the four types of CAPTCHAs, differentiated by age, Internet experience, and education level. The response times are collected for each user. Then, association rules are extracted from user data, for evaluating the dependence of the response time in solving the CAPTCHA from age, education level and experience in internet usage by statistical analysis. The results implicitly capture the users' psychological states showing in what states the users are more sensible. It reveals to be a novelty and a meaningful analysis in the state-of-the-art.
\end{@twocolumnfalse}
\vspace{1cm}

\section{Introduction}
HCI (Human Computer Interaction) studies the interaction between humans, i.e. computer users and computers as well as the major phenomena surrounding them.  Essentially, HCI concerns development, evaluation and implementation of computer systems that are interactive in their core. The interactive design is linked with the feelings, words and their expressions. It represents usability elements that are firmly connected to the user-interface and human factors. Hence, it is deeply involved with the computer science, artificial intelligence and cognitive psychology. From all aforementioned, HCI includes the elements of understanding the computer user needs as well as the usability elements. To be efficient it should be fluent, cognitive, expressive and communicative.

Usability is the main concept in HCI. It is linked with the following elements: (i) easy to learn, (ii) easy to remember how to use, (iii) effective to use, and (iv) efficient to use. From this point of view, the puzzles like CAPTCHA, which humans can easily solve, i.e. decrypt in contrast to the computers, represent an element of HCI. CAPTCHA is an acronym for Completely Automated Public Turing test to tell Computers and Humans Apart. It represents a test program with a task. If the user gives the right answer to the task that the program asks, then the program classifies it as a human, i.e. computer user \cite{ahn90}. Hence, it is created to differentiate the computer users from bots (computer program) in order to effectively login into a certain website. The aim of CAPTCHA is to stop the attacks made by bots. Hence, its main function is to: (i)  protect web sites, applications, interfaces and services like Google, Yahoo or similar, (ii) prevent spam in blogs, and (iii) protect email addresses \cite{google91}.

CAPTCHA involves three different elements: (i) usability, (ii) security, and (iii) practicality \cite{baecher92}. Usability is linked with the process of solving CAPTCHA by the computer users. It measures the efficiency of solving the CAPTCHA as well as the response time computer user needs to solve the CAPTCHA. However, it is sometimes linked with the user's acceptance of the CAPTCHA, too. Security is deeply connected to the difficulty of finding solutions to CAPTCHA by computer. Practicality refers to the easiness of the programmers realization of CAPTCHA.  

In this study, the image-based type of the CAPTCHA will be considered. Among all types of CAPTCHA, the image-based ones are the most related to the elements of HCI, because they are linked to cognitive psychology. Also, the tested CAPTCHA samples are deeply connected to facial expressions, whose solving is influenced by human factors \cite{erickson}. The main goal of this paper is exploring the usability elements like user response time, usage, and preferences related to the image-based CAPTCHA which includes the facial expressions. We used four different image-based CAPTCHAs for testing purposes. Each of them consists of five different images and only one is related to the solution of CAPTCHA. In the experiment, the given CAPTCHAs are tested on a population of 100 different Internet users. The tested Internet users are differentiated by the given attributes: (i) age, (ii) Internet experience, and (iii) educational level. Specifically, we investigate (i) users' response time relating to the usability aspects of the CAPTCHA, and (ii) usage statistics connected to the response time of efficiently solving image-based CAPTCHA which includes elements of facial expression. To process tested data, the association rules are used. In this way, the user response time to efficiently solve certain CAPTCHA is analyzed by association rule method. The obtained results will show the natural dependence and relationship between different variables which are of importance to the way of solving the CAPTCHA. Also, this approach identifies the influence of facial expression to efficiently solve the image-based CAPTCHA. Hence, it is invaluable for choosing a certain type of CAPTCHA to be accustomed to a certain group of Internet users. At the end, it improves the knowledge about the symbiotic system, because it gives information about to what human states the users are more sensible, determining a speed response in solving the CAPTCHA.

The paper is organized as follows. Section 2 describes the image-based CAPT-CHA.  Section 3 clarifies the data mining elements like association rules, which will be used as the main evaluation tool. Section 4 describes the elements of the experiment. Section 5 presents the results and gives the discussion. Section 6 draws conclusions.

\section{CAPTCHA}
\begin{figure}[!h]
\begin{center}
\subfigure{
\includegraphics[width=7cm, height=7cm, keepaspectratio]{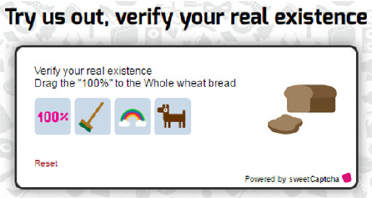}
}
\subfigure{
\includegraphics[width=7.3cm, height=7.3cm, keepaspectratio]{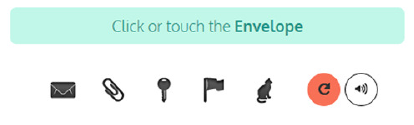}
}
\caption{Image-based CAPTCHA types}
\label{fig1}
\end{center}
\end{figure}

All CAPTCHAs can be divided into the following groups: (i) text-based CAPTCHA, (ii) video based CAPTCHA, (iii) audio-based CAPTCHA, and (iv) image-based CAPTCHA. Text-based CAPTCHA is the  most commonly used CAPTCHA. It usually includes distorted text with different background. To solve this type of CAPTCHA, the distorted characters should be correctly recognized. It is easy to generate,  but easily exposed to vulnerable attacks from bot due to efficient OCR software. Video-based CAPTCHA uses video to create a puzzle to be solved. YouTube videos are typically used as a basis. To be solved, a CAPTCHA user is asked to tag videos with descriptive keywords. The users can reach an efficiency of solving CAPTCHA up to 90\%. Audio-based CAPTCHA usually plays a set of characters or words. The users should recognize and type them in the designated area.  This type of CAPTCHA is critically exposed to bot attacks due to rapid development in the speech recognition. The attack rate can reach up to 70\%. Image-based CAPTCHA is usually considered as the most advanced and safest one. This type of CAPTCHA requires users to find a desired image between the list of images. Because it is based on image details, it represents a very difficult task for bots. Furthermore, this type of CAPTCHA can easily integrate HCI elements connected to cognitive psychology. Fig. \ref{fig1} shows some image-based CAPTCHA types.

Existing image-based CAPTCHAs generally rely on image classification. In that way, the users see a series of images, which have to be identified as well as the relationship between them. Identifying the facial expression is a task accustomed to the humans \cite{erickson}. Hence, it is easy to be solved by humans, but impossible task for bots. Accordingly, integrating such HCI elements in an image-based CAPTCHA can create a very successful CAPTCHA in terms of security, usability as well as practicality.

\section{Association Rule Mining}
Association Rules (ARs) are the most intuitive tool to find the frequent sets of feature values and to understand the co-occurrence of these values in a dataset \cite{agrawal93}.
We start by defining the set of items representing  all values $n$ features can assume as $T=T_1\cup T_2\:\cup,...,\cup\: T_n$, where $T_i=\{t_i^1,t_i^2,...,t_i^x\}$ is the set of $x$ items which are the possible values of feature $i$.
Accordingly, a transaction $I$ in our case is a subset of $n$ items from $T$, such that each item is a value of a distinct feature. This means that
$\forall t_i^j, t_k^h\in I$, $i \ne k$, and $j$ and $h$ are two possible values respectively of feature $i$ and $k$. From the aforementioned, an Association Rule (AR) is defined as an implication $W  \Rightarrow Z$, characterized by \emph{support} and \emph{confidence}, where $W$ and $Z$ represent disjoint sets of items, called respectively antecedent and consequent.
Support quantifies the frequency of $W\cup Z$ in the transactions inside the dataset.
It is related to the statistical significance of AR.
Confidence measures the frequency of $W\cup Z$, given the transactions containing $W$. It evaluates the conditional probability of $Z$ given $W$.

The main goal is the extraction of the ARs whose support value is $\ge minsupport$ and whose confidence value is $\ge minconfidence$, by adopting the well-known \emph{Apriori} algorithm \cite{agrawal94}. It is composed of two main phases: (i) finding all the sets of items with frequency $\ge$ \emph{minsupport} in the dataset; (ii) determining the ARs from obtained sets of items, taking into account the \emph{minconfidence} threshold.
The concept underlying the algorithm is based on the \emph{anti-monotonicity} property: if a set of items is unfrequent, also every its superset will be. Accordingly, unfrequent sets of items are eliminated in each iteration from the algorithm. First of all, the algorithm finds the sets of items of size $k$ whose frequency is $\ge minsupport$. Then, the sets of items are enlarged to size $k+1$, by inserting the only sets of items of size 1 and frequency $\ge minsupport$. These two phases are iterated from $k=1$ to a given value of $k$, such that sets of items of size $k+1$ with frequency $\ge minsupport$ cannot be detected further. Secondly, rules are created from each frequent set of items $F$. It is performed by detecting all the subsets $f \subset F$ such that confidence value of $f \to F - f$ is $\ge minconfidence$.

Another interesting measure to evaluate the ARs has been introduced in \cite{Brin97}, and it is called \emph{lift}.
Lift quantifies the number of times $W$ and $Z$ co-occur more often than expected if they were statistically independent \cite{Hahsler2015}. If the lift assumes a value of 1, then $W$ and $Z$ are independent.
On the contrary, if the lift assumes a value greater than 1, $W$ and $Z$ will co-occur more often than expected. It indicates that the occurrence of $W$ has a positive effect on the occurrence of $Z$.

\section{Experiment}
\begin{figure}[!h]
\begin{center}
\subfigure{
\includegraphics[width=9cm, height=9cm, keepaspectratio]{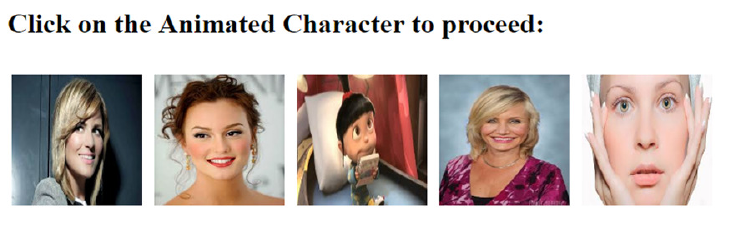}
}
\subfigure{
\includegraphics[width=9cm, height=9cm, keepaspectratio]{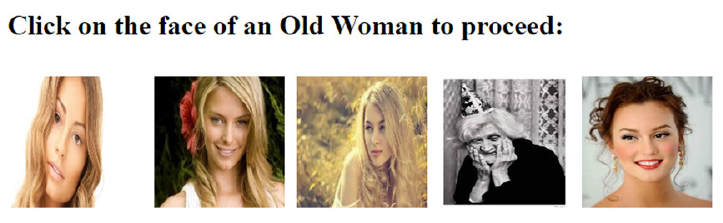}
}
\subfigure{
\includegraphics[width=9cm, height=9cm, keepaspectratio]{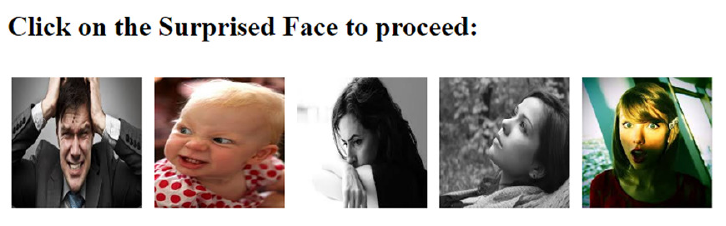}
}
\subfigure{
\includegraphics[width=9cm, height=9cm, keepaspectratio]{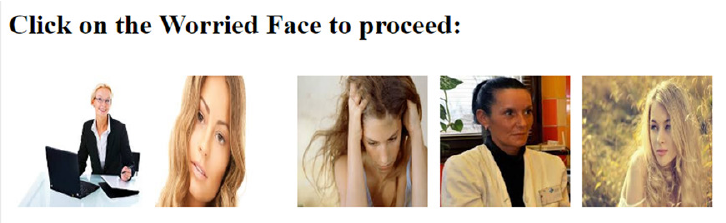}
}

\caption{Image-based CAPTCHAs incorporating facial expressions: animated character, old woman, surprised face and worried face}
\label{fig2}
\end{center}
\end{figure}

Our experiment measures the usability elements of the proposed image-based CAPTCHAs, which incorporate facial expression elements. In that sense, the time needed to efficiently solve a  certain CAPTCHA is measured. Each CAPT-CHA includes five images representing a certain facial expression. One of the images presents the asked facial expression among the other four images which are not. Hence, the users should choose the right image among the five given under consideration. The main goal is to measure and differentiate the influence of facial expression to efficiently solve the image-based CAPTCHA linked with cognitive elements. We propose four CAPTCHAs representing the following facial expressions: (i) animated character, (ii) old woman, (iii) surprised face, and (iv) worried face. Fig. \ref{fig2} shows the proposed CAPTCHAs.

\subsection{Participants}
An experiment is conducted on a population sample of 100 Internet users. The first half of them is composed of women, and the other half is composed of men. The age of participants is from 18 to 52 years old. They have a secondary or higher education level. The population sample includes students, engineers, officials, but also unemployed people. Participants have an Internet experience of at least one up to nine years. All users have joined the experiment voluntarily, knowing that their data would be made publicly available for study purposes, but without publishing their names (anonymous poll). The interview of each participant was conducted personally, by asking to solve the image-based CAPTCHAs. In particular, each Internet user is required to solve the 4 different types of the aforementioned image-based CAPTCHAs. For each user, the response times (in seconds) for finding the solution to the CAPTCHAs are registered. Hence, participants are differentiated by their: (i) age, (b) educational level, and (iii) working experience on the Internet. Also, they are characterized by a certain response time in solving the image-based CAPTCHAs.

\subsection{Dataset Creation}
At the end, a dataset including 7 features (age, education level, number of years of internet usage, response time in solving the animated character, old woman, surprised face, and worried face CAPTCHAs) is created, for a total of 100 instances of 7 features each.
The feature corresponding to the age of the Internet users is linearly divided into two groups, which are below 35 years and above 35 years.
The education level feature can assume also two values, which are the higher education and the secondary education levels.
About the number of years of Internet usage, this feature has integer values from 0 to $+\infty$. In particular, in the dataset it varies from 1 to 9. Consequently, it is discretized by employing an Equal-Width Discretization \cite{discr}, obtaining 3 different ranges, named as high Internet usage ($>$ 6 years), middle Internet usage (from 4 to 6 years), and low Internet usage ($<$ 4 years).
The features corresponding to the response time in solving the 4 different types of image-based CAPTCHA have decimal values varying from 0.0 to $+ \infty$. A test has been conducted for discretization of these features with multiple methods, including Equal-Width Discretization. It demonstrated that most of the methods fail in this task because they obtain a poor discretization. On the contrary, discretization by adopting the K-Medians clustering algorithm \cite{kmedians}, whose advantage is finding natural groups of data based on their characteristics, performed successfully, identifying 3 different ranges.
The aim of the K-Medians is to determine $K$ ranges from the values of the features, by minimizing a function $J$. It measures the total sum of $L_1$ norm between each value in a certain range and its centroid, which is its median value.
We run the K-Medians algorithm 10 times with $K=3$ and selected the set of ranges with the lowest value of $J$ as the optimal solution.
Hence, K-Medians found 3 ranges corresponding to low response time ($<$ 12.34 s), middle response time (from 12.34 s to 27.30 s), and high response time ($>$ 27.30 s).
Table \ref{tab2} reports the feature values, eventually discretized.
\begin{table}
\caption{Possible values and corresponding ranges for each feature of the dataset}\label{tab2}
{\scriptsize
\begin{center}
\begin{tabular}{r@{\quad}r@{\quad}rl}
\hline
Features&Values&Interval\\
\hline\rule{0pt}{12pt}
age&below 35&-\\
&above 35&- \\\\
education level&higher education&-\\
&secondary education&-\\\\
number of years of Internet usage&high Internet usage& $>$ 6 years\\
&middle Internet usage& 4 - 6 years\\
&low Internet usage& $<$ 4 years\\\\
response time in solving animated character CAPTCHA&high response time& $>$ 27.30 s\\
&middle response time& 12.34 s - 27.30 s\\
&low response time&$<$ 12.34 s\\\\
response time in solving old woman CAPTCHA&high response time& $>$ 27.30 s\\
&middle response time& 12.34 s - 27.30 s\\
&low response time&$<$ 12.34 s\\\\
response time in solving surprised face CAPTCHA&high response time& $>$ 27.30 s\\
&middle response time& 12.34 s - 27.30 s\\
&low response time&$<$ 12.34 s\\\\
response time in solving worried face CAPTCHA&high response time& $>$ 27.30 s\\
&middle response time& 12.34 s - 27.30 s\\
&low response time&$<$ 12.34 s\\\\
\hline
\end{tabular}
\end{center}
}
\end{table}

\subsection{Association Rules Extraction}
The obtained dataset is subjected to the \emph{Apriori} algorithm for detection of the ARs, by which we investigate on the association among the different feature values. Specifically, we find sets of frequently co-occurring feature values and that are related to a given response time in solving the 4 types of image-based CAPTCHA. Also, by tuning the \emph{minconfidence} and \emph{minsupport} to appropriate values, we are able to evaluate the strength of such a relationship. It determines a psychological analysis of the HCI, because we are able to understand (i) to which human conditions (age, education level, number of years of internet usage) is more probable the image-based CAPTCHA to be easily solved, and (ii) what type of image-based CAPTCHA is easier to be solved by a human subject. It is an interesting psychological analysis, if we think that each type of image-based CAPTCHA corresponds to a given condition or "state of mind" (worried, surprised, etc.), to which human subjects could be more or less sensible.
For this reason, analysis is split into 4 parts, each corresponding to detect the ARs for a given type of image-based CAPTCHA.
This means that we extract all the ARs by adopting the \emph{Apriori} algorithm. Then, we filter them by selecting those ARs having only the response time feature value of a certain type of image-based CAPTCHA as consequent. Finally, we group the ARs based on the response time of the different types of image-based CAPTCHA, and evaluate them separately.

In our case, $n$ is equal to 7 features. All the possible values of each of them are reported in Table \ref{tab2}.
An example of transaction $I$, corresponding to a row of the dataset, with eventually discretized feature values, is given in Table \ref{tab1}.
\begin{table}[!ht]
\caption{Example of transaction $I$ (row of the dataset)}\label{tab1}
{\tiny
\begin{tabular}{c|c|c|c|c|c|c}
  \hline
 Age  & Educ. level&Num. years Int. usage&Time anim. char. &Time old wom. &Time surpr. face & Time worr. face\\
  \hline
  above 35 & sec. educ. &middle Int. usage &middle resp. time &middle resp. time&middle resp. time&low resp. time \\
  \hline
   \end{tabular}
   }
\end{table}

Also, an example of AR defined on this transaction is given below:\\\\
\indent\indent\indent\indent\indent (Educ. level) \indent (Num. years Internet usage)  (Resp. time worried face)
\vspace{-0.1cm}
\begin{equation}
\:secondary \: educ., \:middle \: Internet\: usage \to low\: response\: time
\end{equation}

The antecedent $W$ of the AR is \{secondary educ., middle Internet usage\}, while the consequent $Z$ is \{low response time\}. We can observe that $W \cup Z$, which is \{secondary educ., middle Internet usage, low response time\}, is found inside the transaction in Table \ref{tab1}. The meaning of this AR is that a secondary education level and a middle experience in Internet usage are enough to easily solve the worried face CAPTCHA.

\section{Results and Discussion}
Feature discretization, extraction and post-processing of the ARs have been performed in MATLAB R2012a on a notebook quad-core at 2.2 GHz, 16 GB RAM and UNIX operating system.

Tables \ref{tab3}-\ref{tab6} show the ARs, sorted by antecedent, detected from the dataset by the \emph{Apriori} algorithm, whose consequent is the response time in solving the 4 different types of CAPTCHA. For each AR, we report the antecedent $W$, the consequent $Z$, the support, the confidence and the lift values. In order to extract ARs of interest, we fix the \emph{minsupport} and \emph{minconfidence} respectively to 5\% and to 50\%.
The differences between the tables corresponding to missing ARs with the same antecedent are denoted as '-'.
It is worth to note that the lift is always greater than 1. It indicates that the extracted ARs are statistically meaningful and that antecedent and consequent are always positively correlated.

\begin{table}[!b]
\caption{Sorted association rules extracted from the dataset, for which the consequent is the response time in solving animated character CAPTCHA}\label{tab3}
\begin{center}
\begin{tabular}{r@{\quad}r@{\quad}r@{\quad}r@{\quad\quad}rl}
\hline
Antecedent&Consequent&Supp.&Conf.&Lift\\
\hline\rule{0pt}{12pt}
 below 35&	 Low resp. time &	0.50	&1.00&1.47\\
 below 35, Higher educ.& 	 Low resp. time& 	0.24&	1.00&1.47\\
 below 35 Secondary educ.& 	 Low resp. time &	0.26&	1.00&1.47\\
 below 35, Low internet usage &	 Low resp. time &	0.12&	1.00&1.47\\
 below 35, Middle internet usage &	 Low resp. time & 	0.26	&1.00&1.47\\
 below 35, High internet usage &	 Low resp. time &	0.12&	1.00&1.47\\
 below 35, Higher educ., Middle internet usage & 	 Low resp. time &	0.15&	1.00&1.47\\
 below 35, Higher educ., High internet usage & 	 Low resp. time &	0.09&	1.00&1.47\\
 below 35, Secondary educ., Low internet usage &	 Low resp. time &	0.12&	1.00&1.47\\
 below 35, Secondary educ., Middle internet usage & 	 Low resp. time &	0.11&	1.00&1.47\\
 High internet usage &	 Low resp. time &	0.15&	0.88&1.30\\
 Higher educ., High internet usage &	 Low resp. time &	0.12&	0.86&1.26\\
 above 35, Secondary educ., Middle internet usage& 	 Middle resp. time& 	0.08&	0.73&2.51\\
 above 35, Middle internet usage& 	 Middle resp. time &	0.20&	0.71&2.46\\
 above 35, Higher educ., Middle internet usage &	 Middle resp. time& 	0.12&	0.71&2.43\\
 Higher educ. &	 Low resp. time &	0.33&	0.69&1.01\\
 above 35, Higher educ.&	 Middle resp. time &	0.15&	0.63&2.15\\
 above 35& 	 Middle resp. time& 	0.29&	0.58&2.00\\
 above 35, Secondary educ.&	 Middle resp. time& 	0.14&	0.54&1.86\\
\\[2pt]
\hline
\end{tabular}
\end{center}
\end{table}

\begin{table}[!b]
\caption{Sorted association rules extracted from the dataset, for which the consequent is the response time in solving old woman CAPTCHA}\label{tab4}
\begin{center}
\begin{tabular}{r@{\quad}r@{\quad}r@{\quad}r@{\quad\quad}rl}
\hline
Antecedent&Consequent&Supp.&Conf.&Lift\\
\hline\rule{0pt}{12pt}
 below 35&	 Low resp. time &	0.49&	0.98&1.31\\
 below 35, Higher educ. &	 Low resp. time &	0.24	&1.00&1.33\\
 below 35, Secondary educ. &	 Low resp. time &	0.25&	0.96&1.28\\
 below 35, Low internet usage& 	 Low resp. time &	0.11&	0.92&1.22\\
 below 35, Middle internet usage& 	 Low resp. time &	0.26&	1.00&1.33\\
 below 35, High internet usage & 	 Low resp. time& 	0.12&	1.00&1.33\\
 below 35, Higher educ., Middle internet usage &	 Low resp. time& 	0.15	&1.00&1.33\\
 below 35, Higher educ., High internet usage& 	 Low resp. time &	0.09	&1.00&1.33\\
 below 35, Secondary educ., Low internet usage &	 Low resp. time &	0.11&	0.92&1.22\\
 below 35, Secondary educ., Middle internet usage & 	 Low resp. time& 	0.11&	1.00&1.33\\
 High internet usage&	 Low resp. time &	0.16&	0.94&1.25\\
 Higher educ., High internet usage&	 Low resp. time& 	0.13	&0.93&1.24\\
 above 35, Secondary educ., Middle internet usage&	 Middle resp. time &	0.06&	0.55&2.37\\
-&-&&&\\			
-&-&&&\\			
 Higher educ.&	 Low resp. time &	0.40	&0.83&1.11\\
-&-&&&\\			
-&-&&&\\			
 above 35, Secondary educ.& 	 Middle resp. time& 	0.14&	0.54&2.34	\\		
 Higher educ., Middle internet usage& 	 Low resp. time& 	0.26&	0.81&1.08\\
 Middle internet usage& 	 Low resp. time &	0.42&	0.78&1.04\\
 above 35, Secondary educ., Low internet usage& 	 Middle resp. time& 	0.08&	0.53&2.32\\
 above 35, Low internet usage &Middle resp. time &	0.09&	0.53&2.30\\
 \\[2pt]
\hline
\end{tabular}
\end{center}
\end{table}

In particular, we can observe that animated character CAPTCHA is the easiest to be solved among the 4 types of CAPTCHA (see Table \ref{tab3}). 
It is mainly because in general the ARs with a low response time exhibit higher support and confidence values than in the other types of CAPTCHA. It indicates that a low response time is likely to occur more often than in the other types of CAPTCHA.
In fact, users below 35 years are associated to a low response time, also independently from the education level, which has no meaningful impact in solving this CAPTCHA, and from the level of internet usage, with a support of 0.50, a confidence of 1.00 and a lift of 1.47.
The second more difficult type of CAPTCHA to solve is the old woman one (see Table \ref{tab4}).  In fact, the pattern for the users below 35 years is maintained with similar but slightly lower values of support and confidence with respect to animated character CAPTCHA. Because in general the support and confidence values of the rules whose consequent is the low response time are smaller in the case of old woman than in the case of animated character, we will have a smaller probability to determine a low response time in solving the CAPTCHA.
 On the other hand, the ARs extracted for old woman CAPTCHA present the highest number of differences with respect to the other types of CAPTCHA (see the four missing rules denoted as '-' in Table \ref{tab4}).
The third type of CAPTCHA more difficult to solve is the surprised face one (see Table \ref{tab5}), having the same pattern for users below 35 years, but corresponding lower support and confidence values, indicating that correlation between users below 35 years and low response time is weaker than before.
Also, a higher education level is sufficient to obtain a low response time in animated character, while it is able to solve the surprised face CAPTCHA only in a middle response time.
The last type of CAPTCHA presenting the highest difficulty level is the worried face one (see Table \ref{tab6}). We can observe that the extracted ARs for this CAPTCHA are very similar to the ARs extracted for surprised face CAPTCHA. In particular, the pattern for the users below 35 years is also visible, but with support and confidence values less than in the surprised face CAPTCHA, and in general in all the other types of CAPTCHA. It indicates that a smaller percentage of transactions contains the co-occurrence of users below 35 years and low response time. This means that users below 35 years are less correlated to a low response time in solving this CAPTCHA. In general, the differences in the 4 types of CAPTCHA represent distinctive characteristics of each type of CAPTCHA that is mandatory to deeply analyze.

\begin{table}[!b]
\caption{Sorted association rules extracted from the dataset, for which the consequent is the response time in solving surprised face CAPTCHA}\label{tab5}
\begin{center}
\begin{tabular}{r@{\quad}r@{\quad}r@{\quad}r@{\quad\quad}rl}
\hline
Antecedent&Consequent&Supp.&Conf.&Lift\\
\hline\rule{0pt}{12pt}
  below 35&	 Low resp. time &	0.44&	0.88&1.49\\
 below 35, Higher educ.& 	 Low resp. time &	0.20&	0.83&1.41\\
 below 35, Secondary educ.& 	 Low resp. time& 	0.24&	0.92&1.56\\
 below 35, Low internet usage& 	 Low resp. time &	0.10&	0.83&1.41\\
 below 35, Middle internet usage& 	 Low resp. time &	0.23&	0.88&1.50\\
 below 35, High internet usage& 	 Low resp. time &	0.11&	0.92&1.55\\
 above 35, Higher educ., Middle internet usage& 	 Middle resp. time& 	0.15&	0.88&2.32\\
 below 35, Higher educ., High internet usage& 	 Low resp. time& 	0.08&	0.89&1.51\\
 below 35, Secondary educ., Low internet usage& 	 Low resp. time& 	0.10&	0.83&1.41\\
 below 35, Secondary educ., Middle internet usage& 	 Low resp. time& 	0.11&	1.00&1.69\\
 High internet usage& 	 Low resp. time &	0.13&	0.76&1.30\\
 Higher educ., High internet usage& 	 Low resp. time& 	0.10&	0.71&1.22\\
 above 35, Secondary educ., Middle internet usage& Middle resp. time& 	0.06&	0.55&1.43\\
 above 35, Middle internet usage& 	 Middle resp. time& 	0.21&	0.75&1.97\\
 below 35, Higher educ., Middle internet usage &Low resp. time& 	0.12&	0.80&1.35\\
 Higher educ.& 	 Middle resp. time &	0.24&	0.50&1.31\\
 above 35, Higher educ.&	 Middle resp. time &	0.20&	0.83&2.19\\
 above 35& 	 Middle resp. time& 	0.32	&0.64&1.68\\
-&-&&&\\						
 Secondary educ., Middle internet usage&	 Low resp. time &	0.16&	0.73&1.23\\
 Secondary educ.& 	 Low resp. time &	0.35&	0.67&1.14\\
 Secondary educ., Low internet usage& 	 Low resp. time& 	0.16&	0.59&1.00\\
 Higher educ., Middle internet usage&	 Middle resp. time& 	0.18&	0.56&1.48\\
  \\[2pt]
\hline
\end{tabular}
\end{center}
\end{table}

Specifically, looking at Table \ref{tab3}, we can observe that users above 35 years and with a middle internet usage are associated to a middle response time in solving the animated character CAPTCHA, also independently from the education level. This pattern is not present for the old woman CAPTCHA (see Table \ref{tab4}), for which users above 35 years and with a middle internet usage are associated to a middle response time only in the case of secondary education level. Furthermore, in animated character CAPTCHA, we find that users above 35 years are related to a middle response time, also independently from education level (hence, also with higher education level) and number of years of Internet usage. On the contrary, for old woman CAPTCHA (see Table \ref{tab4}), the strongest rule in terms of support, confidence and lift  (respectively of 0.14, 0.54 and 2.34) involving the users above 35 years indicates that users above 35 years are able to solve the CAPTCHA in a middle time if they have a secondary education level.
Again, looking at Table \ref{tab5} of surprised face CAPTCHA, another difference can be observed with animated character CAPTCHA. In particular, it is required a higher education level to the users above 35 years to solve this CAPTCHA in a middle time. On the contrary, for the animated character CAPTCHA, users above 35 years are able to solve it in a middle time also when they have a secondary education level.
In the case of worried face CAPTCHA in Table \ref{tab6}, we observe a meaningful difference with animated character CAPTCHA for the users above 35 years. In fact, in the case of animated character CAPTCHA, users above 35 years are associated to a middle response time also if they have a secondary education level, while, in the case of worried face, if they have a higher education level, with similar values of support, confidence and lift. A last difference is visible between the ARs of the surprised face CAPTCHA and those of the worried face CAPTCHA. In fact, in the second ones, a middle internet usage is related to a middle response time in solving that CAPTCHA, with a support of 0.28, a confidence of 0.52 and a lift of 1.18. Differently, this rule is not present for the surprised face CAPTCHA.

Finally, we can make the following observations: (i) Internet users below 35 years have a good attitude in solving the image-based CAPTCHA; (ii) the easiest type of image-based CAPTCHA to solve is the animated character one, followed by old woman, surprised face and the more complex worried face ones; (iii) users above 35 years are able to solve the animated character CAPTCHA in a moderate time; (iv) users above 35 years have an attitude to solve the old woman CAPTCHA in a moderate time if they have a secondary education level; (v) it is required a higher education level to the users above 35 years to solve the surprised face and the worried face CAPTCHAs in a moderate time.

\begin{table}[!ht]
\caption{Sorted association rules extracted from the dataset, for which the consequent is the response time in solving worried face CAPTCHA}\label{tab6}
\begin{center}
\begin{tabular}{r@{\quad}r@{\quad}r@{\quad}r@{\quad\quad}rl}
\hline
Antecedent&Consequent&Supp.&Conf.&Lift\\
\hline\rule{0pt}{12pt}
  below 35& 	 Low resp. time& 	0.38	&0.76&1.43\\
 above 35, Higher educ.&	 Middle resp. time& 	0.21&	0.88&1.99\\
 below 35, Secondary educ.&	 Low resp. time &	0.23	& 0.88&1.67\\
 below 35, Low internet usage& 	 Low resp. time &	0.09&	0.75&1.41\\
 below 35, Middle internet usage& 	 Low resp. time& 	0.19&	0.73&1.38\\
 below 35, High internet usage& 	 Low resp. time &	0.10&	0.83&1.57\\
 above 35, Higher educ., Middle internet usage& 	 Middle resp. time& 	0.16&	0.94&2.14\\
 below 35, Higher educ., High internet usage&	 Low resp. time& 	0.07&	0.78&1.47\\
 below 35, Secondary educ., Low internet usage&	 Low resp. time &	0.09&	0.75&1.41\\
 below 35, Secondary educ., Middle internet usage& 	 Low resp. time &	0.11&	1.00&1.89\\
 High internet usage& 	 Low resp. time& 	0.12&	0.71&1.33\\
 Higher educ., High internet usage& 	 Low resp. time& 	0.09&	0.64&1.21\\
 above 35, Secondary educ., Middle internet usage& 	 Low resp. time &	0.06&	0.55&1.03\\
 above 35, Middle internet usage& 	 Middle resp. time& 	0.21&	0.75&1.70\\
 below 35, Higher educ., Middle internet usage,&	 Low resp. time &	0.08&	0.53&1.01\\
 Higher educ.& 	 Middle resp. time &	0.30	&0.63&1.42\\
 below 35, Higher educ.& 	 Low resp. time &	0.15&	0.63&1.18\\
 above 35& 	 Middle resp. time &	0.32	&0.64&1.45\\
-&-&&&&\\					
 Secondary educ., Middle internet usage& 	 Low resp. time &	0.17&	0.77&1.46\\
 Higher educ., Middle internet usage& 	 Middle resp. time& 	0.23&	0.72&1.63\\
 Secondary educ.& 	 Low resp. time &	0.35&	0.67&1.27\\
 Secondary educ., Low internet usage& 	 Low resp. time &	0.15&	0.56&1.05\\
 Middle internet usage& Middle resp. time& 	0.28&	0.52&1.18 \\
  \\[2pt]
\hline
\end{tabular}
\end{center}
\end{table}

\section{Conclusions}
The paper analyzed the human response to the facial expression given in specific image-based CAPTCHA. Hence, it explored the computer users' response time (reaction rate) to the given CAPTCHA according to cognitive psychology linked with facial expression recognition. The obtained results were statistically processed by the ARs. The analysis by ARs revealed interesting co-occurrences of feature values and their association with a given response time to solve the facial expression image-based CAPTCHA. Although the results were quite similar, the AR methodology proved to be powerful to differentiate subtle variations between processed values. In that way, the results showed that computer users can more easily recognize CAPTCHAs that incorporate animated characters. Also, the facial expressions like surprised or worried face are the most difficult to recognize and differentiate. It is worth to note that recognition of the old woman among the other images was characterized with different elements to be influenced to than the images presented in other analyzed CAPTCHAs. The obtained results and further research in the given area can help in finding the most appropriate CAPTCHA for a wider range of Internet users. Also, the knowledge about the symbiotic system is improved by showing in what human states the users are more sensible.

In this context, ARs have not been properly used for prediction, but for unsupervised detection of the natural dependencies and relationships between feature values and response times. Hence, future work will be focused on decision tree-based methods for prediction of the response times by supervised learning. Analysis will be further enriched by providing a larger dataset and a literature review of cognitive science on how people differentiated by age or educational level react to different image stimuli. It will be useful to formalize hypothesis on the expected correlations between the response time and the other features. 

\subsection*{Acknowledgments.}
The authors are grateful to the participants for publicly providing their data. %

\end{document}